\documentclass{sig-alternate}
\pdfoutput=1
\usepackage{listings,amsmath, amssymb}
\usepackage{times}
\usepackage{xspace}
\usepackage{float}
\usepackage{color}
\usepackage{graphicx}
\begin{document}

%%%%%%%%%%%%%%%%%%%%%%%%%%%%%%%%%%
%%  compacting tricks           %%
%%%%%%%%%%%%%%%%%%%%%%%%%%%%%%%%%%
% Compact itemize and enumerate.  Note that they use the same counters and
% symbols as the usual itemize and enumerate environments.
\def\compactify{\itemsep=0pt \topsep=0pt \partopsep=0pt \parsep=0pt}
\let\latexusecounter=\usecounter
\newenvironment{CompactEnumerate}
  {\def\usecounter{\compactify\latexusecounter}
   \begin{enumerate}}
  {\end{enumerate}\let\usecounter=\latexusecounter}
\newenvironment{CompactItemize}
  {\def\usecounter{\compactify\latexusecounter}
   \begin{itemize}}
  {\end{itemize}\let\usecounter=\latexusecounter}
\newenvironment{Proof}{\par \noindent{\bf Proof.}}{\(\QED\) \par}
\newcommand{\QED}{\mbox{}\hspace*{\fill}\nolinebreak\mbox{$\Box$}}
%%%%%%%%%%%%%%%%%%%%%%%%%%%%%%%%%%
%%     Defining thrms, etc.     %%
%%%%%%%%%%%%%%%%%%%%%%%%%%%%%%%%%%
\newtheorem{theorem}{{Theorem}}
\newtheorem{definition}{Definition}
\newtheorem{lemma}{{Lemma}}
\newtheorem{fact}{{Fact}}
\newtheorem{corollary}{{Corollary}}
\newtheorem{proposition}{{Proposition}}
\newtheorem{conjecture}{{Conjecture}}
\newtheorem{claim}{{Claim}}
%%%%%%%%%%%%%%%%%%%%%%%%%%%%%%%%%%
%%   commands to ignore text    %%
%%%%%%%%%%%%%%%%%%%%%%%%%%%%%%%%%%
\newcommand{\comment}[1]{}
\newcommand{\ignore}[1]{}
\newcommand{\delete}[1]{}
%%%%%%%%%%%%%%%%%%%%%%%%%%%%%%%%%%
%%     typefacing reminders     %%
%%%%%%%%%%%%%%%%%%%%%%%%%%%%%%%%%%
\newcommand{\calE}{ {\cal E}}
\newcommand{\scword}{ {\sc Word}}
%%%%%%%%%%%%%%%%%%%%%%%%%%%%%%%%%%
%%       other reminders        %%
%%%%%%%%%%%%%%%%%%%%%%%%%%%%%%%%%%
\newcommand{\fbycase}[5]{
#1
\left\{ \begin{array}{ll}
    {#2}&{\mbox{#3}}\\
    {#4}&{\mbox{#5}}
    \end{array}
\right.
}

\newcommand{\stack}[2]{
\begin{array}{c}
    {#1}\\{#2}
\end{array}
}
%%%%%%%%%%%%%%%%%%%%%%%%%%%%%%%%%%

\title{SHALE: An Efficient Algorithm for Allocation of Guaranteed Display Advertising}

\numberofauthors{8}

\author{
\alignauthor
Vijay Bharadwaj \\
    \affaddr{Netflix}\thanks{Work done while at Yahoo!Labs}\\
    \email{bharadway.vijay@gmail.com}
\alignauthor
Peiji Chen\\
    \affaddr{Yahoo!Labs}\\
    \email{peiji@yahoo-inc.com}
\alignauthor
Wenjing Ma\\
    \affaddr{Yahoo!Labs}\\
    \email{wenjingm@yahoo-inc.com}
\and
\alignauthor
Chandrashekhar Nagarajan\\
    \affaddr{Yahoo!Labs}\\
    \email{cn54@yahoo-inc.com}
\alignauthor
John Tomlin\\
    \affaddr{opTomax Solutions}\thanks{Work done while at Yahoo!Labs}\\
    \email{johntomlin@acm.org}
\alignauthor
Sergei Vassilvitskii\\
    \affaddr{Yahoo!Labs}\\
    \email{sergei@yahoo-inc.com}
\and
\alignauthor
Erik Vee\\
    \affaddr{Yahoo!Labs}\\
    \email{erikvee@yahoo-inc.com}
\alignauthor
Jian Yang\\
    \affaddr{Yahoo!Labs}\\
    \email{jianyang@yahoo-inc.com}
}

\maketitle

%%%%%%%%%%%%%%%%%%%%%%%%%%%%%%%%%%
%%        macros                %%
%%%%%%%%%%%%%%%%%%%%%%%%%%%%%%%%%%
\newcommand{\eps}{\varepsilon}
\newcommand{\elg}{\sim}
\newcommand{\elig}{\sim}
\newcommand{\neij}{\Gamma(j)}
\newcommand{\neii}{\Gamma(i)}
\newcommand{\g}{g_{ij}}
\newcommand{\rem}{\tilde{s}}

\newcommand{\at}[1][t]{\alpha^{#1}}
\newcommand{\bt}[1][t]{\beta^{#1}}

\begin{abstract}
Motivated by the problem of optimizing allocation 
 in guaranteed display advertising,
we develop an efficient, lightweight method of generating a compact {\em allocation plan} that
can be used to guide ad server decisions.
The plan itself uses just $O(1)$ state per guaranteed contract, is robust to noise, and allows
us to serve (provably) nearly optimally.
The optimization method we develop is scalable, with a small in-memory footprint, and working
in linear time per iteration.  It is also ``stop-anytime,'' meaning that time-critical applications
can stop early and still get a good serving solution.
Thus, it is particularly useful for optimizing the large problems arising in the context
of display advertising.
We demonstrate the effectiveness of our algorithm
using actual Yahoo! data. 
\end{abstract}

\section{Introduction}\label{sec:intro}

A key problem in display advertising is how to efficiently
serve in some (nearly) optimal way.
As internet publishers and advertisers become increasingly sophisticated,
it is not enough to simply
make serving choices ``correctly'' or ``acceptably''.  
Improving objective goals by just a few percent can often improve revenue by tens of millions of dollars
for publishers, as well as improving advertiser or user experience.
Serving needs to be done in such a way that we maximize
the potential for users, advertisers, and publishers.

In this paper, we address serving display advertising
in the guaranteed display marketplace, providing a lightweight optimization framework
that allows real servers to allocate ads efficiently and with little overhead.
Recall that in guaranteed display advertising, advertisers may target particular
types of users visiting particular types of sites over a specified time period.
Publishers guarantee to serve their ad some promised number of times to users matching the
advertiser's criteria over the specified duration.  We refer to this as a {\em contract}.

In~\cite{vvs10}, the authors show that given a forecast of future inventory, 
it is possible to create an optimal {\em  allocation plan},
which consists of labeling each contract with just $O(1)$ additional information.
Since it is so compact, this allocation plan can efficiently be communicated to ad servers.  It 
requires no online state, which removes the need for maintaining immediately accessible impression counts.
(An {\em impression} is generated whenever there is an opportunity to display an ad somewhere on a web
page for a user.)
Given the plan, each ad server can easily decide which ad to serve each impression, even when the impression
is one that the forecast never predicted.  The delivery produced by following the plan is nearly optimal.
Note that simply using an optimizer to find an optimal allocation of contracts to impressions would not produce
such a result, since the solution is too large and does not generalize to unpredicted outputs.

The method to generate the allocation plan outlined in~\cite{vvs10} relies on the ability to
solve large, non-linear optimization problems; it takes as input a bipartite graph representing
the set of contracts and a sample of predicted user visits, which can have hundreds of millions of arcs or more. 
There are commercially available solvers
that can be used to create allocation plans.  However, they have several drawbacks.  The most
prominent of these is that such solvers aim towards finding good primal solutions, while the
allocation plan generated is not directly tied to the quality of such solutions.  (The allocation plan relies on
the dual solution of the problem.)  In particular,
there is no guarantee of how close to optimal  the allocation plan really is.
Hence, although creating a good allocation plan is time critical, stopping
the optimizer early with sub-optimal values can have undesirable effects for serving.

For our particular problem, the graph
we wish to optimize is extremely large and scalability becomes a real concern.
For this reason, and given the other disadvantages of using complex third party software, 
we propose a new solution, called  `SHALE.'  It addresses all of these concerns, having many
desirable properties:
\begin{itemize}
\item
It has the ``stop anytime'' property.  That is, after completing any iteration, we can stop SHALE and produce a good answer.
\item
It is a multi-pass streaming algorithm.  Each iteration of SHALE runs as a streaming algorithm, reading the arcs off disk one at a time.
The total online memory is proportional to the number of contracts and samples used, and is independent of
the number of edges in the graph.  Because of this, it is possible to handle inputs that are prohibitively
large for many commercial solvers without special modifications.
\item
It is guaranteed to converge to the true optimal solution if it runs for enough iterations.  It is robust to sampling, so
the input can be generated by sampling rather than using a full input.  
\item
Each contract is annotated with just $O(1)$ information,
which can be used to produce nearly optimal serving.  Thus, the solution
generated creates a practical allocation plan, useable in real serving systems.
\end{itemize}
The SHALE solver uses the idea of~\cite{vvs10} as a starting point, but it provides an additional
twist that allows the solver to stop after any number of iterations and still produce a good allocation
plan.  For this reason, SHALE is often five times faster than solving the full problem using
a commercial solver.

\subsection{Related Work}\label{sec:related}
The allocation problem facing a display advertising publisher has been the subject of increased attention in the past few years.  Often modeled as a special version of a stochastic optimization, several theoretical solutions have been developed ~\cite{stochastic2, vahab}. 
A similar formulation of the problem was done by Devanur and Hayes~\cite{Devanur},who added an assumption that user arrivals are drawn independently and identically from some distribution, and then proceed to develop allocation plans based on the learned distribution.  In contrast, Vee et al.~\cite{vvs10} did not assume independence of arrivals, but require the knowledge of the user distributions to formulate the optimization problem. 

Bridging the gap between theory and practice, Feldman et al.~\cite{stochastic1} demonstrated that primal-dual methods can be effective for solving the allocation problem. However, it is not clear how to scale their algorithm to instances on billions of nodes and tens of billions of edges.  A different approach was given by  Chen et al.~\cite{microsoftAlloc} who used the {\em structure} of the allocation problem to develop control theory based methods to guide the online allocation and mitigate the impact of potential forecast errors. 

Finally, a crucial piece of all of the above allocation problems is the underlying optimization function. Ghosh et al~\cite{repbid} define representative allocations, which minimize the average $\ell_2^2$ distance between an allocation given to a specific advertiser, and the ideal one which allocates every eligible impression with equal probability.  Feldman et al.~\cite{stochastic1} define a similar notion of {\em fair} allocations, which attempt to minimize an $\ell_1$ distance between the achieved allocation and a similarly defined ideal.

\section{Problem Statement}\label{sec:preliminaries}
In this section, 
we begin by defining the notion of an optimal allocation of ads to users/impressions
(Section~\ref{sec:optimal allocation}).
Our goal will then be to serve as close as possible to this optimal allocation.
In Section~\ref{sec:compact serving}, we describe the notion of generating an allocation
plan, which will be used to produce nearly optimal serving.

\subsection{Optimal Allocation}
\label{sec:optimal allocation}
\begin{figure}
{\centering
\includegraphics[scale=0.4]{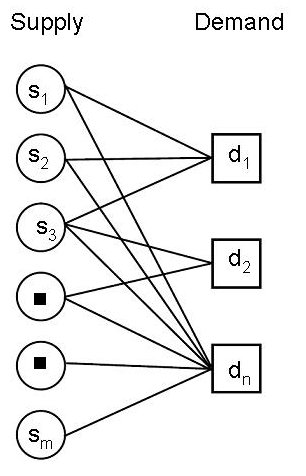}
\par}
\caption{Example bipartite graph} \label{fig:bipartite1}
\end{figure}

In guaranteed display advertising, we have a large number of forecast impressions together with a
number of contracts.  
These contracts specify a {\em demand} as well as a target; we must deliver a number of impressions
at least as large as the specified demand, and further, each impression must match the target specified by the contract. 
We model this as a bipartite graph.  On one side are {\em supply nodes}, representing impressions. 
On the other side are {\em demand nodes}, representing contracts.  We add an arc from a given supply node to a given demand node 
if and only if the impression that the supply node represents is {\em eligible} (i.e. matches the target profile) 
for the contract represented by the demand node.  Further, demand nodes are labeled with a {\em demand}, which is precisely the 
amount of impressions guaranteed to the represented
contract.  In general, supply nodes will represent several impressions each, thus each supply node is labeled with a weight $s_i$, leading to a weighted graph  (see~\cite{vvs10} for more details).
Figure~\ref{fig:bipartite1} shows a simple example. 

An optimal allocation must both be feasible and minimize some objective function.  Here, our
objective balances two goals: minimizing penalty, and maximizing {\em representativeness}.  Each demand node/contract $j$
has an associated penalty, $p_j$.  
Let $u_j$ be the {\em under-delivery}, i.e. the number of impressions delivered less than $d_j$.
Then our total penalty is $\sum_j p_j u_j$.

Representativeness is a measure of how close our allocation is to some target.  For each impression $i$ and contract $j$,
we define a target, $\theta_{ij}$.  In this paper, we set $\theta_{ij} = d_j / S_j$, where $S_j = \sum_{i\in\neij} s_i$,
the total eligible supply for contract $j$.  This has the effect of aiming for an equal mix of all possible matching impressions. (Here, $\neij$ is the neighborhood of $j$, likewise, we denote the neighborhood
of $i$ by $\neii$.)
The non-representativeness for contract $j$ is the weighted $L_2$ distance from the target $\theta_{ij}$ and the proposed
allocation, $x_{ij}$.  Specifically, 
$$
    \frac{1}{2}\sum_{i\in\neij} s_i\frac{V_j}{\theta_{ij}}(x_{ij} - \theta_{ij})^2 ,
$$
where $V_j$ is the relative priority of the contract $j$; a larger $V_j$ means that representativeness is more important.
Notice that we weight by $s_i$ to account for the fact that some sample impressions have more weight than others.  
% Dividing by \theta_{ij}?
Representativeness is key for advertiser satisfaction.  Simply
giving an advertiser the least desirable type of users 
(say, three-year-olds with a history of not spending money) 
or attempting
to serve out an entire contract in a few hours decreases long-term revenue by driving advertisers away.
See~\cite{repbid} for more discussion on this idea.

Given these goals, we may write our optimal allocation in terms of a convex optimization problem:
\begin{eqnarray}
\nonumber
\mbox{Minimize\ \ \ } & \frac{1}{2} \sum_{j, i\in\neij} s_i\frac{V_j}{\theta_{ij}}(x_{ij} - \theta_{ij})^2
                                + &\sum_{j} p_j u_j \\
\label{eq:demand}
\mbox{s.t.}&\sum_{i\in\neij} s_i x_{ij} + u_j \ge d_j & \forall j \\
\label{eq:supply}
&\sum_{j\in\neii} x_{ij} \le 1 & \forall i \\
\label{eq:nonneg}
& x_{ij}, u_j \ge 0 & \forall i,j
\end{eqnarray}

Constraints~\ref{eq:demand} are called {\em demand constraints}.  They guarantee that 
$u_j$ precisely represents the total underdelivery to contract $j$. 
Constraints~\ref{eq:supply} are {\em supply constraints}, and they specify that we serve no more than one ad for each
impression.  Constraints~\ref{eq:nonneg} are simply {\em non-negativity constraints}.

The {\em optimal allocation} for the guaranteed display ad problem is the solution to the above problem, where
the input bipartite graph represents the full set of contracts and the {\em full set of impressions!}
Of course, generating the full set of impressions is impossible in practice.  The work of~\cite{vvs10}
shows that using a sample of impressions still produces an approximately optimal fractional allocation.
We interpret the fractions as the {\em probabilities} that a given impression should be allocated to a given contract. Since there are billions of impressions, this  leads to serving that is nearly identical.

Although this paper focuses on the above problem, we note that our techniques can be extended to more general
objectives.  For example, in related work,~\cite{clickpaper} 
described a multi-objective model for the allocation of inventory to guaranteed delivery, which combined
penalties and representativeness (as above) with revenue made on the non-guaranteed display (NGD) spot market
and the potential revenue gained from supplying clicks to contracts.  SHALE can easily be extended to handle
these variants.  

\subsection{Compact Serving}
\label{sec:compact serving}
In the previous subsection, we defined the notion of optimal allocation.  However, serving such an allocation
is itself a different problem.  Following~\cite{vvs10}, we define the problem of online serving with forecasts as follows.

We are given as input a bipartite graph, as described in the previous subsection.  (We assume this graph is an
approximation of the future inventory, although it is not necessary for this definition.)
We proceed in two phases.

\begin{itemize}
\item {\bf Offline Phase}:  Given the bipartite graph as input, we must annotate each demand node (corresponding to a contract) 
with $O(1)$ information.  This information will guide the allocation during the online phase.

\item {\bf Online Phase}:
During the online phase, impressions arrive one at a time.  For each impression, we are given 
the set of eligible contracts, together with the annotation computed during the offline phase of each returned contract.
Using only this information, we must decide which contract to serve to the impression.
\end{itemize}
The {\em online allocation} is the actual allocation of impressions to contracts given during the 
online phase.  Our goal is to produce an online allocation that is as close to optimal as possible.

Remarkably, the work of [1] shows that there is an algorithm that solves the above problem nearly optimally.
If the input bipartite graph exactly models the future impressions, then the online allocation produced is optimal.
If the input bipartite graph is generated by sampling from the future, then the online allocation produced is
provably approximately optimal.

However, the previous work simply assumed that an optimal solution can be found during the Offline Phase.
Although this is true, it does not address many of the practical concerns that come with solving large-scale
non-linear optimization problems.  In the following sections, we describe our solution, which in addition
to solving the problem of compact serving, is fast, simple, and robust.

\section{Algorithms}\label{sec:solution}

\subsection{Plan creation using full solution}
The proposal of~\cite{vvs10} to create an allocation plan was to solve the
problem of Section~\ref{sec:optimal allocation} using standard methods.  From this,
we can compute the {\em duals} of the problem.  In particular,
we may write the problem in terms of its Lagrangian (more formally, we use the KKT conditions).
Every constraint then has a corresponding dual variable.  (Intuitively, the harder a constraint
is to satisfy, the larger its dual variable in the optimal solution.)

The allocation plan then consists of the demand duals of the problem, denoted $\alpha$.  So
each contract $j$ was labeled with the demand dual from the corresponding demand constraint, $\alpha_j$.
The supply duals, denoted $\beta$, and the non-negativity duals were simply thrown out.

A key insight of this earlier work is that we can reconstruct the optimal solution using only the
$\alpha$ values.
When impression $i$ arrives, the value of $\beta_i$ can be found online by solving the equation
$\sum_{j\in\neii} g_{ij}(\alpha_j - \beta_i) = 1$, resetting $\beta_i = 0$ if the solution is less than 0.
Here, $g_{ij}(z) = \max\{0 , \theta_{ij}(1 + z/V_j)\}$.
We then set $x_{ij} = g_{ij}(\alpha_j - \beta_i)$ for each $j\in\neii$.  Somewhat surprisingly, this yields
an optimal allocation.  (And when the value of $\alpha$ is obtained by solving a sampled problem, it
is approximately optimal.)

As mentioned in the introduction, although this solution has many nice properties, 
solving the optimization problem using standard methods is
slower than desirable.  Thus, we have a need for faster methods.

\subsection{Greedy solution (HWM)}

An alternate approach to solving the allocation problem is the {\em High Water Mark} (HWM) algorithm, based on a greedy heuristic.
This method first orders all the contracts by their {\em allocation order}.  Here, the allocation order puts contracts
with smaller $S_j$ (i.e. total eligible supply) before contracts with larger $S_j$.
Then, the algorithm goes through
each contract one after another, trying to allocate an equal fraction from all the eligible ad opportunities.
This fraction is denoted $\zeta$ for each contract, and corresponds roughly to its demand dual.
Contract $j$ is given fraction $\zeta_j$ from each eligible impression, {\em unless} previous contracts have taken
more than a $1-\zeta_j$ fraction already.  In this case, contract $j$ gets whatever fraction is left (possibly 0).

If there is very little contention (or contract $j$ comes early in the allocation order), then $\zeta_j = d_j/S_j$.
This will give exactly the right amount of inventory to contract $j$.  However, if a lot of inventory has already been allocated
when $j$ is processed, its $\zeta_j$ value may be larger than this to accommodate the fact that it gets less 
than $\zeta_j$ for some impressions.  Setting $\zeta = 1$ will give a contract all inventory that has not already been allocated.
We do this in the case that there is not enough remaining inventory to satisfy the demand of $j$.

The pseudo-code is summarized as follows.

\begin{enumerate}
\item Order all demand nodes in decreasing contention order ($d_j/S_j$).
\item For each supply node $i$,
    initialize the available weight $\tilde{s_i}=s_i$.
\item For each demand node $j$, in allocation order:
\label{alg:hwm step}
   \begin{enumerate}
    \item Find $\zeta_j$ such that 
    \label{alg:hwm substep}
        \[\sum_{i \in B_j} \min \{\tilde{s_i}, \zeta_j s_i\} = d_j, \]
    setting $\zeta_j = \infty$ if the above has no solution. 
     \item For each matching supply nodes $i \in B_j$\\
        Update $\tilde{s_i} = \tilde{s_i} - \min \{\tilde{s_i}, \zeta_j s_i\}$.
    \end{enumerate}
\end{enumerate}
We note that the computation in Step~\ref{alg:hwm substep} can be
done in time linear in the size of $|B_j|$.  Hence, the total runtime of the HWM
algorithm is linear in the number of arcs in the graph.

\subsection{SHALE}        
Obtaining a full solution using traditional methods is too slow (and more precise than needed), while
the HWM heuristic, although very fast, sacrifices optimality.  SHALE is a method that spans the two approaches.
If it runs for enough iterations, it produces the true optimal solution.  Running it for 0 iterations (plus an additional step
at the end) produces the HWM allocation.  So we can easily balance precision with running time.  In our experience 
(see Section \ref{sec:experiments}), just
10 or 20 iterations of SHALE yield remarkably good results; for serving, even using 5 iterations works quite well since
forecast errors and other issues generally dwarf small variations in the solution.  
Further, SHALE is amenable to ``warm-starts,'' using
the previous allocation plan as a starting point.  In this case, it is even better.

SHALE is based on the solution using optimal duals.  The key innovation, however, is the ability to take
{\em any} dual solution and convert it into a good primal solution.  We do this by extending the simple heuristic
HWM to incorporate dual values.  Thus, the SHALE algorithm has two pieces.  The first piece finds reasonable
duals.  This piece is an iterative algorithm. On each iteration, the dual solution will generally improve.  (And repeated
iterations converge to the true optimal.)
The second piece converts the reasonable set of duals we found (more precisely, the $\alpha$ values, as described earlier)
into a good primal solution.

The optimization for SHALE relies heavily on the machinery provided by the KKT conditions.  Interested
readers may find a more detailed discussion in the Appendix.  Here, we note the following.
If $\alpha^*$ and $\beta^*$ are optimal dual values, then
\begin{enumerate}
\item
The optimal primal solution is given by $x^*_{ij} = \g(\alpha^*_j - \beta^*_i)$, where
$\g(z) =  \max\{0 , \theta_{ij}(1 + z/V_j)\}$.
\item
For all $j$, $0\leq \alpha^*_j \leq p_j$.  
Further,  either $\alpha^*_j = p_j$ or $\sum_{i\in\neij} s_i x^*_{ij} = d_j$.
\item
For all $i$, $\beta_i \ge 0$.  Further, either $\beta_i = 0$ or $\sum_{j\in\neii} x^*_{ij} = 1$.
\end{enumerate}
The pseudo-code for SHALE is shown below.
\begin{itemize}
\item {\bf Initialize}.  Set $\alpha_j = 0$ for all $j$.
\item {\bf Stage One}. Repeat until we run out of time:
    \begin{enumerate}
    \item For each impression $i$, find $\beta_i$ that satisfies
      \label{alg:SHALE beta}
    \[
        \sum_{j\in\neii} \g(\alpha_j - \beta_i) = 1
    \]
    If $\beta_i < 0$ or no solution exists, update $\beta_i = 0$.
    \item For each contract $j$, find $\alpha_j$ that satisfies
      \label{alg:SHALE alpha}
    \[
        \sum_{i\in\neij} s_i \g(\alpha_j - \beta_i) = d_j
    \]
    If $\alpha_j\!\! >\! p_j$ or no solution exists, update $\alpha_j$ = $p_j$.
    \end{enumerate}
\item {\bf Stage Two}.
    \begin{enumerate}
    \item Initialize $\rem_i = 1$ for all $i$.
    \item For each impression $i$, find $\beta_i$ that satisfies
    \[
        \sum_{j\in\neii} \g(\alpha_j - \beta_i) = 1
    \]
    If $\beta_i < 0$ or no solution exists, update $\beta_i = 0$.
    \item For each contract $j$, in allocation order, do:
        \begin{enumerate}
        \item Find $\zeta_j$ that satisfies
            \[
                \sum_{i\in\neij} \min\{ \rem_i , s_i\g(\zeta_j - \beta_i) \} = d_j ,
            \]
            setting $\zeta_j = \infty$ if there is no solution.
        \item For each impression $i$ eligible for $j$, update $\rem_i = \rem_i - \min\{ \rem_i , s_i\g(\zeta_j -\beta_i)\}$.
        \end{enumerate}
    \end{enumerate}
\item {\bf Output} The $\alpha_j$ and $\zeta_j$ values for each $j$.
\end{itemize}
    
Our implementation of SHALE runs in linear time (in the number of arcs in the input graph) per iteration.
    
During Stage One, we iteratively improve the $\alpha$ values by assuming that the $\beta$ values are correct and solving the
equation
for $\alpha$.  Recall that $x_{ij} = \g(\alpha_j - \beta_i)$.  Thus, we are simply solving the equation
$\sum_{i\in\neij} s_i x_{ij} = d_j$ for $\alpha_j$.  In order to find better $\beta$ values, we assume the $\alpha$ is correct
and solve for $\beta$ using $\sum_{j\in\neii} x_{ij} = 1$.  
The following theorem shows that this simple iterative technique converges, and yields an $\eps$ approximation
in polynomial steps.

More precisely, define $d_j(\alpha) = \sum_{i\in\neij} s_i \g(\alpha_j - \beta_i)$, where $\beta$ is
determined as in Step~\ref{alg:SHALE beta} of Stage One of SHALE.  (We think of this as the 
projected delivery for contract $j$ using only Stage One of SHALE.)
We say a given $\alpha$ solution produces an {\em $\eps$-approximate delivery}
if for all $j$, either $\alpha_j = p_j$ or $d_j(\alpha) \ge (1-\eps)d_j$.
Note that an optimal $\alpha_j$ is at most $p_j$; the intuitive reason for this is that growing $\alpha_j$ any
larger will cause the non-representativeness of the contract's delivery to be even more costly than the under-delivery
penalty.  Thus, an $\eps$-approximate delivery means that every contract is projected to deliver within
$\eps$ of the desired amount, or its $\alpha_j$ is ``maxed-out.''

We can now state our theorem.  Its proof is in the appendix.
\begin{theorem}
\label{thrm:iterations}
Stage One of SHALE converges to the optimal solution of the guaranteed display allocation problem.
Further,
let $\eps>0$.  Then within $\frac{1}{\eps} n \max_j\{ p_j/V_j \}$ iterations,
the output $\alpha$ produces an $\eps$-approximate delivery.
\end{theorem}

Note that Stage One is effectively a form of coordinate descent.  In general, it could be replaced with any standard
optimization technique that allows us to recover a set of approximate dual values.  However, the form we use is
simple to understand, use, and debug.  Further, it works very well in practice.

In Stage Two, we calculate $\zeta$ values in a way similar to HWM.  We calculate $\beta$ values based on the $\alpha$ values
generated from Stage One.  Using these, we calculate $\zeta$ values to give $d_j$ allocation (if possible) to each contract.
Notice that in Stage Two, we must be cognizant of the actual allocation.  Thus, we maintain a remaining fraction left,
$\rem_i$, that we cannot exceed.  Thus, contracts allocated latest may not be able to get the full amount specified
by $\g$, if the fraction taken from impression $i$ is too great.

We note that in our actual implementation, we use a two-pass version of Stage Two.  In the first pass, we bound
$\zeta_j$ by $\alpha_j$ for each $j$.  In the second pass, we find a second set of $\zeta$ values (with no upper bounds), 
utilizing any left-over inventory.  This is somewhat ``truer'' to the allocation produced by SHALE in Stage One, and gives
slightly better online allocation.

\subsubsection{Online Serving with SHALE}
\label{subsec:online_serving}
Recall that SHALE produces two values for each contract $j$, namely $\alpha_j$ and $\zeta_j$.  Given impression $i$, 
the $\alpha$ values for eligible contracts
are used to calculate the $\beta_i$ value, which is used together with the $\zeta$ values to produce the allocation.
The pseudo-code is below.

\begin{CompactEnumerate}
\item[]{\bf Input:} Impression $i$ and the set of eligible contracts.
\item Set $\rem_i = 1$ and find $\beta_i$ such that
\[
    \sum_{j\in\neii} \g(\alpha_j - \beta_i) = 1
\]
If $\beta_i < 0$ or no solution exists, set $\beta_i = 0$.
\item
For each matching contract $j$, in allocation order, 
    compute $x_{ij} = \min\{ \rem_i , \g(\zeta_i - \beta_i)\}$
    and update $\rem_i \leftarrow \rem_i - x_{ij}$.
\item Select contract $j$ with probability $x_{ij}$.  (If $\sum_{j\neii} x_{ij} < 1$, then
there is some chance that no contract is selected.)
\end{CompactEnumerate}

\section{Experiments}\label{sec:experiments}
We have implemented both the HWM and SHALE  algorithms described in
Section~\ref{sec:solution} and benchmarked their performance against
the full solution approach (known hereafter as XPRESS) on
historical booked contract sets. 
We have extensively tuned the parameters for XPRESS, so it is much
faster than just using it ``off-the-shelf.''
First we describe these datasets
and our chosen performance metrics and then present our evaluation
results.

\subsection{Experimental setup}
\label{experimental setup}
In order to test the ``real-world'' performance of all three
algorithms we considered 6 sets of real GD contracts booked and
active in the recent past. In particular, we chose three periods of
time, each for one to two weeks, 
and two ad positions
LREC and SKY for each of these time periods.

We considered US region contracts booked to the aforementioned
positions and time periods and also excluded all frequency capped
contracts and all contracts with time-of-day and other custom targets.
Also, all remaining contracts that were
active for longer than the specified date ranges were truncated and
their demands were proportionally reduced. Next, 
we generated a bipartite graph for each contract set as in
Figure~\ref{fig:bipartite1}; by sampling $50$ eligible impressions for each
contract in the set. This sampling procedure is described in detail in
\cite{vvs10}. We then ran HWM, SHALE and XPRESS on each of the 6 graphs
and evaluated the following metrics.
\begin{enumerate}
\item  {\bf Under-delivery Rate} : This represents the total
  under-delivered impressions as a proportion of the booked demand,
  i.e.,
\begin{equation}
\label{eq:under-delivery rate}
U \ = \ \frac{\sum_{j} u_j}{\sum_{j} d_j}
\end{equation}
\item {\bf Penalty Cost} : This represents the penalty incurred
  by the publisher for failing to deliver the guaranteed number of
  impressions to booked GD contracts. Note that the true long-term penalty 
  due to under-delivery  is not known since we cannot easily forecast
  how an advertiser's future business with the publisher will change due to
  under-delivery on a booked contract. Here we define the total
  penalty cost to be
\begin{equation}
\label{eq: penalty cost def}
P \ = \ \sum_{j}p_ju_j
\end{equation}
where $u_j$ is the number of under-delivered impressions to contract
$j$ and $p_j$ is the cost for each under-delivered
impression. For our experiments, we set $p_j$ to be $p_j =
0.005 \ + \ q_j$ where $q_j$ is the revenue per delivered
impression from contract $j$. Indeed, it is intuitive and reasonable
to expect that contracts that are more valuable to the advertiser
incur larger penalties for under-delivery. The offset (here $\$5 CPM$)
serves to ensure that our algorithms attempt to fully deliver even the
contracts with low booking prices.

\item {\bf L2 Distance} : This metric shows how much the generated
  allocation deviates from a desired allocation (for example a
  perfectly representative one). In particular, the L2 distance is
  the non-representativeness function 
  $\frac{1}{2}\sum_{i\in\neij} s_i\frac{V_j}{\theta_{ij}}(x_{ij} - \theta_{ij})^2$,
  the first term of the objective
  function in Section~\ref{sec:preliminaries}, corresponding to the
weighted $\ell_2^2$ distance between target and allocation.
\end{enumerate}

\subsection{Experiment 1}
\label{subsec:expt 1}
As we mentioned earlier, SHALE was designed to provide a trade-off
between the speed of execution of HWM and the quality of solutions
output by XPRESS. Accordingly in our first experiment we measured the
performance of SHALE (run for 0, 5, 10, 20 and 50 iterations)
as compared to XPRESS against our chosen metrics.  Since SHALE at 0 iterations
is the same as HWM, we label it as such.
\begin{figure}[h!]
\centering
\includegraphics*[viewport=37 155 760 575,scale=0.32]{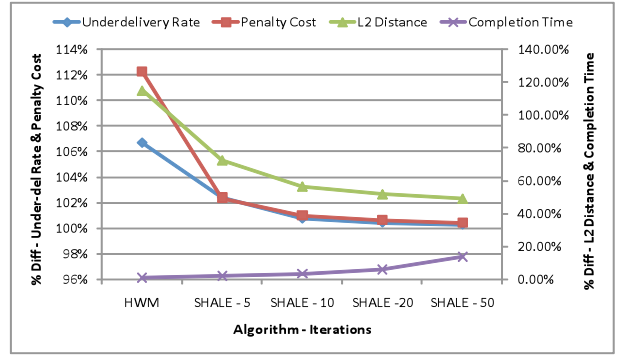}
\caption{Performance Vs. Completion time}
\label{fig:expt 1}
\end{figure}
Figure~\ref{fig:expt 1} shows the penalty cost, under-delivery rate,
L2 distance and completion  for HWM and SHALE run for 5, 10, 20 and 50
iterations respectively as a percentage of the corresponding metric
for XPRESS, averaged over our 6 chosen contract sets. 
Note that the y-axis labels for the under-delivery
rate and penalty cost are on the left, while the labels for the L2 distance and completion time are on the right.

It is immediately clear that SHALE after only 10 iterations is within  2\% of XPRESS with respect to penalty cost and
under-delivery rate. Further, note that SHALE after 10 iterations is able to provide an
allocation whose L2 distance is less than half that of
XPRESS.  (Recall smaller L2 distance means the solution is more representative, so SHALE is doing twice as well on this metric.) 
This somewhat surprising result seems to be an artifact of the SHALE algorithm: The functional form of $\g$ is determined 
by the representativeness objective, so we can think of representativeness as ``driving'' the algorithm.

Even at 50 iterations, SHALE is more than 5 times as fast as XPRESS.  Remarkably, its penalty and under-delivery
are almost equal to XPRESS (less than 1\% different), yet the L2 distance is still much better.
At 20 iterations, we see SHALE gives a very high-quality solution, despite being about an order of
magnitude faster than the commercial solver.

\subsection{Experiment 2}
\label{subsection:expt_serv}
We next study how SHALE performs compared to the optimal 
algorithm when used to serve real world sampled impressions from 
actual server logs. This experiment uses real contracts and real adserver logs 
(downsampled) for performing the complete offline simulation.

\subsubsection{Setup}
Here we take three new datasets which consists of real guaranteed delivery contracts 
from Yahoo! active during different one to two week periods in the past year. We 
run our optimization algorithms and serve real downsampled serving logs for 
each of the one-to-two week periods, reoptimizing every two hours.  That is,
the offline optimizer creates an allocation plan to serve the contracts for the remaining
duration; we serve for two hours using that plan; collect the delivery stats so far;
then re-optimize for the rest of the duration using the updated stats.  Note that the
two-hours corresponds to two hours of serving logs.  Our actual simulation is somewhat
faster due to the downsampling.

\subsubsection{Algorithms compared}
At the end of the simulation, we look at the contracts that start and end within 
the simulation period and compare how metrics of under-delivery and penalty across 
HWM, SHALE and DUAL algorithms.  
Our DUAL solution is obtained by running a coordinate gradient descent algorithm 
till convergencence; if our forecasts had been perfect, this would have produced optimal
delivery.
The SHALE algorithms are run with setting of 0, 5, 10 and 20 iterations, with the 0-iteration
version labeled as HWM.

We performed serving using the reconstruction algorithm described in Section
\ref{subsec:online_serving}. 

\subsubsection{Metrics}
The metrics include the underdelivery metric and penalty metrics as defined in 
Equation \ref{eq:under-delivery rate} and in Equation \ref{eq: penalty cost def}
For these set of experiments, we set $p_j$ to be $p_j =
0.002 \ + \ 4*q_j$ where $q_j$ is the revenue per delivered
impression from contract $j$. 

We also compare another metric called {\em pacing} between these algorithms. This captures 
how representative contracts are with respect to time during the delivery of these contracts.
The {\em linear goal} of a contract at a given time is the amount of delivery was perfectly
smooth with respect to time.  For example, a 7 day contract with demand of 14 million has a linear
goal of 6 milion on day 3.
In this experiment, pacing is defined as the percentage of contracts that are within 12\% of the linear 
delivery goal at least 80\% of their active duration. 

\subsubsection{Results}
\begin{figure}[h!]
\centering
\includegraphics*[scale=0.32]{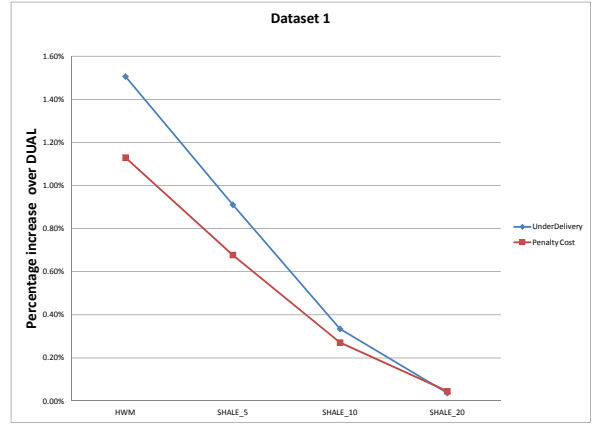}
\caption{Dataset 1: Under Delivery and Penalty Cost Comparison}
\label{fig:dataset1}
\end{figure}

\begin{figure}[h!]
\centering
\includegraphics*[scale=0.32]{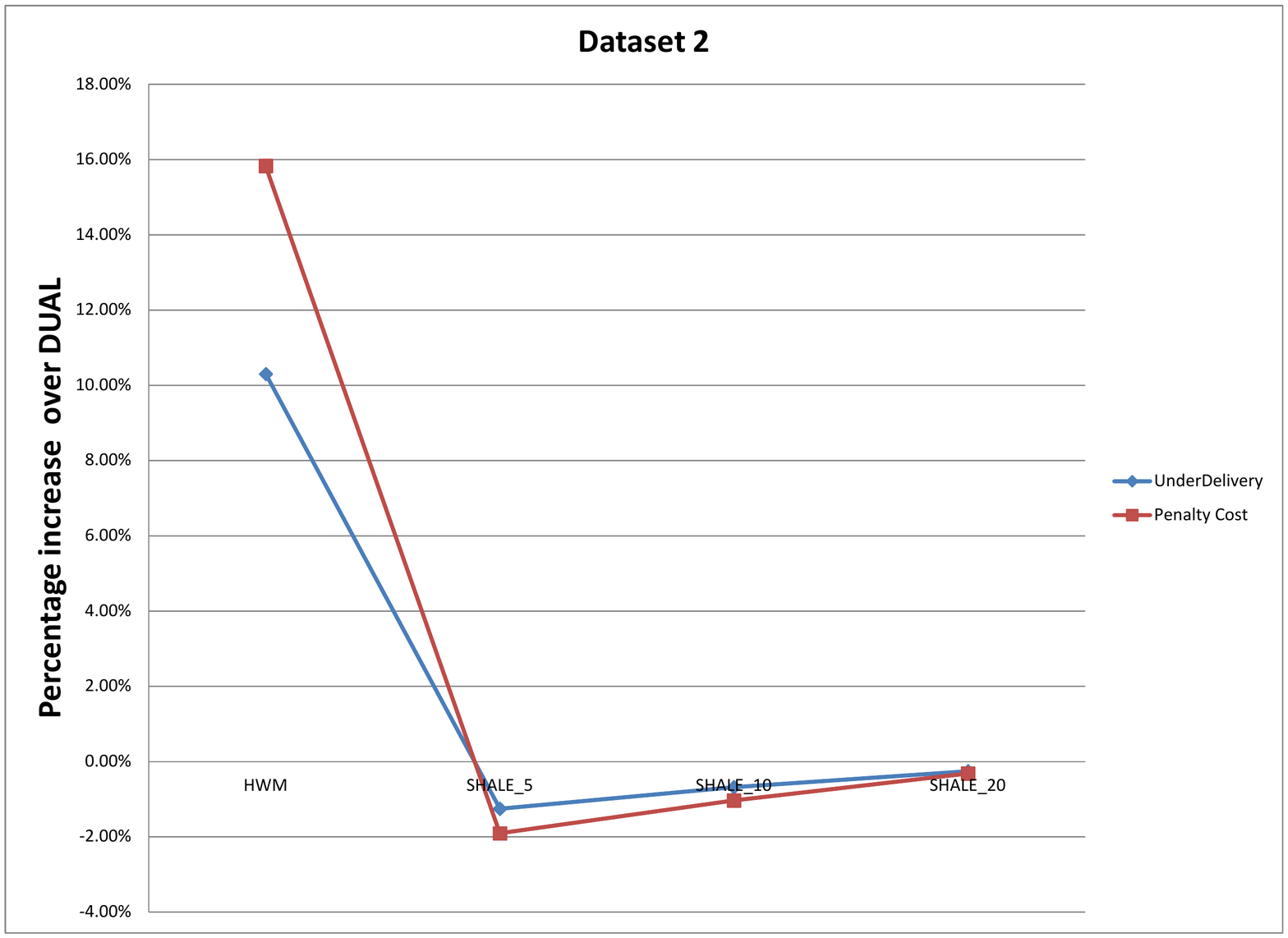}
\caption{Dataset 2: Under Delivery and Penalty Cost Comparison}
\label{fig:dataset2}
\end{figure}

\begin{figure}[h!]
\centering
\includegraphics*[scale=0.32]{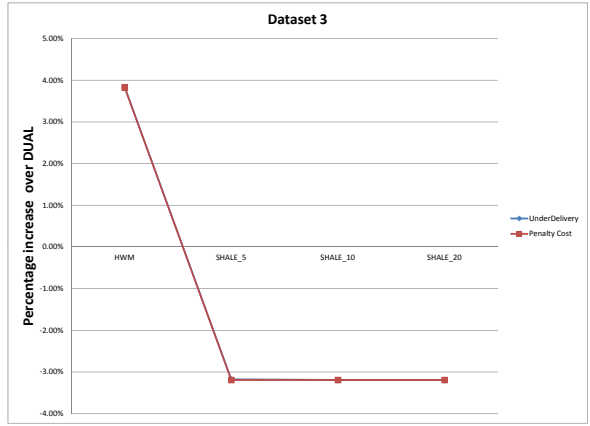}
\caption{Dataset 3: Under Delivery and Penalty Cost Comparison}
\label{fig:dataset3}
\end{figure}

Figures \ref{fig:dataset1}, \ref{fig:dataset2} and \ref{fig:dataset3} show that the 
under-delivery and penalty cost for HWM (SHALE with 0 iterations) algorithm 
is the worst. Further, as the number of SHALE iterations increase it gets very close to the DUAL
algorithm. Note that even SHALE with 5 or 10 iterations performs as well or sometimes 
slightly better than the DUAL algorithm. This can be attributed to different reasons;
one being the fact that there are forecasting errors intrinsic to using real serving logs. 
Another contributing factor is the fact that the DUAL algorithm does not
directly optimize for either of these metrics.
In addition, Stage Two
attempts to fulfill the delivery of every contract, even if it is not optimal according to the objective
function.  This heuristic aspect of SHALE actually appears to aid in its performance when judged by
simple metrics like delivery.

\begin{figure}[h!]
\centering
\includegraphics*[scale=0.32]{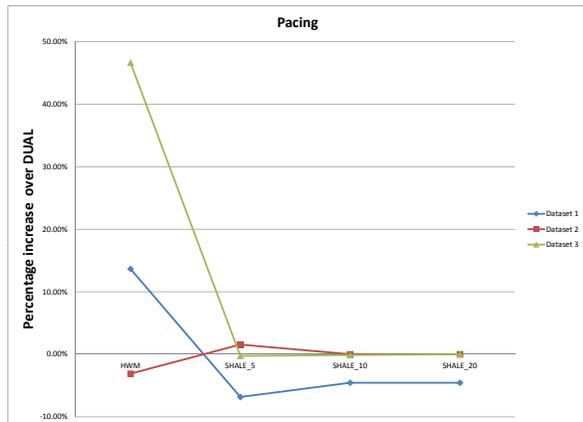}
\caption{Pacing Comparisons on all three datasets}
\label{fig:pacing}
\end{figure}

Figure \ref{fig:pacing} shows how these algorithm perform with respect to pacing.
The pacing is similar for all three datasets for SHALE with 5, 10 and 20 iterations when compared with 
the DUAL algorithm.  Surprisingly, HWM has better pacing than SHALE and DUAL for two of the datasets.
One possible reason for this is that SHALE and DUAL algorithm gives better under-delivery and 
penalty cost, compromising some pacing. Note that the time dimension is just one of the many dimensions
that the representativeness portion of the objective function.
This may also be an artifact of forecasting errors.  In real systems, certain additional modifications are employed
to ensure good pacing.  For these experiments, we have removed those modifications to give a clearer picture of
how the base algorithms perform.

\subsection{Experiment 3}
\label{subsec:expt 2}
\begin{figure}[h!]
\centering
\includegraphics*[viewport=37 155 760 575,scale=0.32]{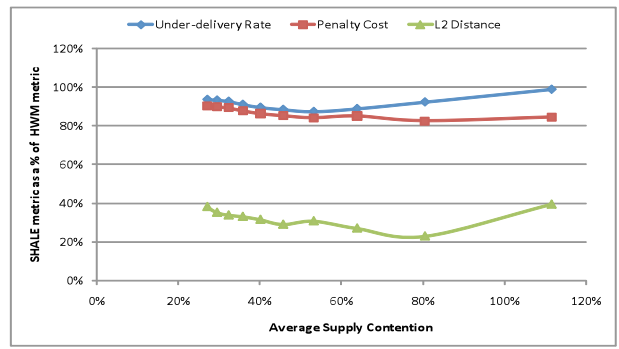}
\caption{SHALE Vs. HWM}
\label{fig:expt 2}
\end{figure}

Superficially, HWM and SHALE both perform well.  In this experiment, we do a  more detailed
simulation to compare HWM and SHALE.
We fix the iteration count for SHALE at 20 and test
its performance under varying supply levels. Specifically, for each of our 6 contract sets, we
artificially reduced the supply weight on each of the supply nodes
while keeping the graph structure fixed in order to simulate the
increasing scarcity of supply. We define the average supply contention (ASC)
metric to represent the scarcity of supply, as follows
\begin{equation}
\label{eqn:avg supply cont}
\textrm{ASC} \ = \ \frac{\sum_i s_i \left(\sum_{j\in
      i}\frac{d_j}{S_j}\right)}{\sum_i s_i}
\end{equation}
where $s_i$ represents the supply weight and $d_j$ and $S_j$ represent
the demand and eligible supply for contract $j$.
In Figure~\ref{fig:expt 2}, we show the
under-delivery rate, penalty cost and L2 distance for SHALE as a
percentage of the corresponding metric for HWM for various levels of
ASC.
First we note that each of our metrics for SHALE is better than the
corresponding metric for HWM for all values of ASC. Indeed, the SHALE
L2 distance is less than 50\% of that for HWM. Also note 
that the SHALE penalty cost consistently improves compared to HWM as
the ASC increases.  This indicates that even though HWM appears to have better
pacing for some data sets, SHALE is still a more robust algorithm and is likely
preferrable in most situations.  (Indeed, we see very consistently that its
under-delivery penalty and revenue are both clearly better.)

\section{Conclusion}
We described the SHALE algorithm, which is used to generate compact allocation plans leading to near-optimal
serving.  Our algorithm
is scalable, efficient, and has the stop-anytime property, making it particularly useful in time-sensitive
applications.  Our experiments demonstrate that it is many times faster than using commercially available general purpose solvers,
while still leading to near-optimal solutions.  On the other side, it produces a much better and more robust solution
than the simple HWM heuristic.  Due to its stop-anytime property, it can be configured to give the desired tradeoff between
running time and optimality of the solution.  Furthermore, SHALE can handle ``warm starts,'' using a previous allocation
plan as a starting point for future iterations.

SHALE is easily modified to handle additional goals, such as maximizing revenue in the non-guaranteed market or 
click-through rate of advertisement.  In fact, the technique appears to be amenable to other
classes of problems involving many users with supply constraints (e.g. each user is shown only one item).   
Thus, although SHALE is particularly well-suited to optimizing guaranteed display ad delivery, it
is also an effective lightweight optimizer.  It can handle huge, memory-intensive inputs, and the underlying techniques
we use provide a useful method of mapping non-optimal dual solutions into nearly optimal primal results.

\pagestyle{empty}
\bibliographystyle{abbrv}
\bibliography{shale}

\begin{thebibliography}{1}

\bibitem{microsoftAlloc}
Y.~Chen, P.~Berkhin, B.~Anderson, and N.~R. Devanur.
\newblock Real-time bidding algorithms for performance-based display ad
  allocation.
\newblock In C.~Apt{\'e}, J.~Ghosh, and P.~Smyth, editors, {\em KDD}, pages
  1307--1315. ACM, 2011.

\bibitem{Devanur}
N.~R. Devenur and T.~P. Hayes.
\newblock The adwords problem: online keyword matching with budgeted bidders
  under random permutations.
\newblock In J.~Chuang, L.~Fortnow, and P.~Pu, editors, {\em ACM Conference on
  Electronic Commerce}, pages 71--78. ACM, 2009.

\bibitem{stochastic1}
J.~Feldman, M.~Henzinger, N.~Korula, V.~S. Mirrokni, and C.~Stein.
\newblock Online stochastic packing applied to display ad allocation.
\newblock In M.~de~Berg and U.~Meyer, editors, {\em ESA (1)}, volume 6346 of
  {\em Lecture Notes in Computer Science}, pages 182--194. Springer, 2010.

\bibitem{stochastic2}
J.~Feldman, A.~Mehta, V.~S. Mirrokni, and S.~Muthukrishnan.
\newblock Online stochastic matching: Beating 1-1/e.
\newblock In {\em FOCS}, pages 117--126. IEEE Computer Society, 2009.

\bibitem{repbid}
A.~Ghosh, P.~McAfee, K.~Papineni, and S.~Vassilvitskii.
\newblock Bidding for representative allocations for display advertising.
\newblock In S.~Leonardi, editor, {\em WINE}, volume 5929 of {\em Lecture Notes
  in Computer Science}, pages 208--219. Springer, 2009.

\bibitem{vahab}
V.~S. Mirrokni, S.~O. Gharan, and M.~Zadimoghaddam.
\newblock Simultaneous approximations for adversarial and stochastic online
  budgeted allocation.
\newblock In D.~Randall, editor, {\em SODA}, pages 1690--1701. SIAM, 2012.

\bibitem{vvs10}
E.~Vee, S.~Vassilvitskii, and J.~Shanmugasundaram.
\newblock Optimal online assignment with forecasts.
\newblock In D.~C. Parkes, C.~Dellarocas, and M.~Tennenholtz, editors, {\em ACM
  Conference on Electronic Commerce}, pages 109--118. ACM, 2010.

\bibitem{clickpaper}
J.~Yang, E.~Vee, S.~Vassilvitskii, J.~Tomlin, J.~Shanmugasundaram,
  T.~Anastasakos, and O.~Kennedy.
\newblock Inventory allocation for online graphical display advertising.
\newblock {\em CoRR}, abs/1008.3551, 2010.

\end{thebibliography}

\section*{Appendix}
\label{sec:appendix}

Recall that our optimization problem is
\begin{eqnarray}
\nonumber
\mbox{Minimize\ \ \ } & \frac{1}{2} \sum_{j, i\in\neij} s_i\frac{V_j}{\theta_{ij}}(x_{ij} - \theta_{ij})^2
                                + &\sum_{j} p_j u_j \\
\mbox{s.t.}&\sum_{i\in\neij} s_i x_{ij} + u_j \ge d_j & \forall j \\
&s_i\sum_{j\in\neii} x_{ij} \le s_i & \forall i \\
& x_{ij}, u_j \ge 0 & \forall i,j
\end{eqnarray}
Notice that we have multiplied the supply constraints by $s_i$ to aid our mathematics later.

The KKT conditions generalize the somewhat more familiar Lagrangian.  Let $\alpha_j$ denote the demand duals.
Let $\beta_i$ denote the supply duals. Let $\gamma_{ij}$ denote the non-negativity duals for $x_{ij}$, and
let $\psi_j$ denote the non-negativity dual for $u_j$.
For our problem, the KKT conditions tell us the optimal primal-dual solution must satisfy the following
\begin{itemize}
\item[] {\bf Stationarity:}
\begin{align*}
&\mbox{For all $i,j$,\ \ } s_i \frac{V_j}{\theta_{ij}}(x_{ij} - \theta_{ij}) - s_i \alpha_j + s_i \beta_i - \gamma_{ij} \\
&\mbox{For all $i$,\ \ }  p_j - \alpha_j - \psi_j = 0 
\end{align*}
\item[]
{\bf Complementary slackness:}
\begin{align*}
&\mbox{For all $j$, either $\alpha_j = 0$ or $\sum_{i\in\neij} s_i x_{ij} + u_j = d_j$.}\\
&\mbox{For all $i$, either $\beta_i = 0$ or $\sum_{j\in\neii} s_i x_{ij} = s_i$.}\\
&\mbox{For all $i,j$, either $\gamma_{ij} = 0$ or $x_{ij} = 0$.} \\
&\mbox{For all $j$, either $\psi_j = 0$ or $u_j = 0$.}
\end{align*}
\end{itemize}
The dual feasibity conditions also tell us that $\alpha_j\ge 0$, $\beta_i\ge 0$, $\gamma_{ij} \ge 0$,
and $\psi_j \ge 0$ for all $i,j$.  (While the primal feasibility conditions tell us that the constraints
in the original problem must be satified.)
Since our objective is convex, and primal-dual solution satisfying the KKT conditions is in fact optimal.

Notice that the stationarity conditions are effectively like taking the derivative of the Lagrangian.  The first of these
tells us that
\[
x_{ij} = \theta_{ij}(1 + \frac{\alpha_j - \beta_i + \gamma_{ij}/s_i}{V_j})
\]
The complementary slackness condition for the $\gamma_{ij}$ tells us that $\gamma_{ij} = 0$ unless $x_{ij} = 0$.  
This has the effect that when the expression $ \theta_{ij}(1 + \frac{\alpha_j - \beta_i}{V_j})$ is negative,
$\gamma_{ij}$ will increase just enough to make $x_{ij} = 0$.  In particular, this implies
\[
x_{ij} = \max\{0, \theta_{ij}(1 + \frac{\alpha_j - \beta_i}{V_j})\} = \g(\alpha_j - \beta_i)
\]
The second stationarity condition shows $\alpha_j = p_j - \psi_j$.  Since $\psi_j \ge 0$, this immediately
shows that $\alpha_j \leq p_j$.  Further, the complementary slackness condition for $\psi_j$ implies
that $\psi = 0$ unless $u_j = 0$.  That is, either $\alpha_j = p_j$ or $\sum_{i\in\neij} s_i x_{ij} \ge d_j$.
By complementary slackness of $\alpha_j$, we see in fact that equality must hold (i.e.
$\sum_{i\in\neij} s_i x_{ij} = d_j$) unless $\alpha_j = 0$.  But when $\alpha_j = 0$,
inspection reveals that $\sum_{i\in\neij} s_i x_{ij} = \sum_{i\in\neij} s_i \g(- \beta_i) \le d_j$.
Hence, even when $\alpha_j = 0$, equality must hold for an optimal $\alpha_j$.

Finally, the complementary slackness condition on $\beta_i$ implies either $\beta_i = 0$
or $\sum_{j\in\neii} x_{ij} = 1$.
Putting all of this together, we see that
\begin{enumerate}
\item
The optimal primal solution is given by $x^*_{ij} = \g(\alpha^*_j - \beta^*_i)$, where
$\g(z) =  \max\{0 , \theta_{ij}(1 + z/V_j)\}$.
\item
For all $j$, $0\leq \alpha^*_j \leq p_j$.  
Further,  either $\alpha^*_j = p_j$ or $\sum_{i\in\neij} s_i x^*_{ij} = d_j$.
\item
For all $i$, $\beta_i \ge 0$.  Further, either $\beta_i = 0$ or $\sum_{j\in\neii} x^*_{ij} = 1$.
\end{enumerate}
as we claimed in Section~\ref{sec:solution}.

\begin{proof}[of Theorem~\ref{thrm:iterations}]
First, note that $\alpha_j$ is bounded above by $p_j$.  We will show that $\alpha_j$ is non-decreasing on
each iteration.  Let $\at$ refer to the value of alpha computed during the $t$-th iteration, where
$\at[0]_j = 0$ for all $j$. 
We show by induction that $d_j(\at) \leq d_j$ for all $t\geq 0 $. 
The base case follows by definition, since $\beta_i \ge 0$ for all $i$:
$d_j(\at[0]) \leq \sum_{i\in\neij} s_i \g(0 - 0) = \sum_{i\in\neij} s_i \theta_{ij} = d_j$.

So assume for some $t\ge 0$ that $d_j(\at) \leq d_j$ for all $j$.  Let $\bt$ be the value
computed in Step~\ref{alg:SHALE beta} of Stage One of SHALE, given $\at$.  
We see that
\begin{align*}
d_j(\at) &= \sum_{i\in\neij} s_i \g(\at[t]_j - \bt_i) \\
    &= \sum_{i\in\neij} s_i \max\{0, \theta_{ij} (1 + \frac{\at[t]_j - \bt_i}{V_j})\} 
\end{align*}
Further, by the way in which $\at[t+1]$ is calculated (in Stage One, Step~\ref{alg:SHALE alpha}),
we have that
$\at[t+1]_j$ must either be $p_j$ or satisfy the following:
\begin{align*}
d_j &= \sum_{i\in\neij} s_i \g(\at[t+1]_j - \bt_i) \\
    &= \sum_{i\in\neij} s_i \max\{0, \theta_{ij} (1 + \frac{\at[t+1]_j - \bt_i}{V_j})\} 
\end{align*}
Using the fact that for any numbers $a\ge b$ that $\max\{0, a\} - \max\{0, b\} \leq a-b$ (which
can be shown by an easy case analysis), we have
\begin{align*}
d_j -d_j(\at) 
    &= \sum_{i\in\neij} s_i \max\{0, \theta_{ij} (1 + \frac{\at[t+1]_j - \bt_i}{V_j})\} \\
     &\ \ \ -  \sum_{i\in\neij} s_i \max\{0, \theta_{ij} (1 + \frac{\at[t]_j - \bt_i}{V_j})\} \\
    &\leq \sum_{i\in\neij} s_i \theta_{ij} (\at[t+1]_j - \at[t]_j)/V_j \\ 
    & = d_j(\at[t+1]_j - \at[t]_j)/V_j
\end{align*}
That is, either $\at[t+1]_j = p_j$ or
\begin{align}
\label{eqn:alpha increase}
\at[t+1]_j = \at_j + V_j (1-\frac{d_j(\at)}{d_j})
\end{align}
Since $d_j(\at) \leq d_j$ by assumption, this shows that $\at[t+1]_j \ge \at_j$ for each $j$.
We must still prove that $d_j(\at[t+1]) \leq d_j$.  To this end, note that the $\bt[t+1]$ generated in
Step~\ref{alg:SHALE beta} for the given $\at[t+1]$ must greater than or equal to $\bt$, since
$\at[t+1] \ge \at$.  That is,
$\bt[t+1]_i \ge \bt_i$ for all $i$.  Thus,
\begin{align*}
d_j(\at[t+1]) 
    &= \sum_{i\in\neij} s_i \max\{0, \theta_{ij} (1 + \frac{\at[t+1]_j - \bt[t+1]_i}{V_j})\} \\
    &\leq \sum_{i\in\neij} s_i \max\{0, \theta_{ij} (1 + \frac{\at[t+1]_j - \bt[t]_i}{V_j})\} \\
    &= d_j
\end{align*}
as we wanted.  

In general, we can use the fact that $d_j(\at) \leq d_j$ for all $t$, together with Equation~\ref{eqn:alpha increase},
to see that the $\alpha_j$ values are non-decreasing at each iteration.  From this (together with the fact that $\alpha_j$ is bounded by
$p_j$), it immediately follows that the algorithm converges.  

To see that the algorithm converges to the optimal
solution, we note that the dual values generated by SHALE satisfy the KKT conditions at convergence:
for all $j$, either $\alpha_j = p_j$ or $d_j(\alpha) = d_j$ (i.e. $p_j - \alpha_j - \psi_j = 0$ with either
$\psi_j = 0$ or $u_j = 0$), with similar arguments holding for the other duals.
Since the problem we study is convex, this shows that the primal solution generated must be the optimal.

As for our second claim, suppose that there is some $j$ for which $\at_j \neq p_j$ 
but $d_j(\at_j) \leq (1-\eps)d_j$.  Then by Equation~\ref{eqn:alpha increase}, we see
\[
\at[t+1]_j = \at_j + V_j(1 - \frac{d_j(\at)}{d_j}) \geq \at_j + V_j \eps
\]
That is, $\at[t+1]_j$ increases (over $\at_j$) by at least $\eps V_j$.  Since
$\at_j$ starts at 0, is bounded by $p_j$, and never decreases, we see
that this can happen at most $p_j / (\eps V_j)$ times for each $j$.
In this worst case, this happens for every $j$, giving us the bound we claim.

\end{proof}

\end{document}